# Identifying OCRs in cfDNA WGS Data by Correlation Clustering

Farshad Noravesh 1 ∗, Fahimeh Palizban 2 ∗

*These authors contributed equally to this work.

1-noraveshfarshad@gmail.com

2- palizbanfahimeh@gmail.com

**Abstract**

In the recent decade, the emergence of liquid biopsy has significantly improved cancer monitoring and detection. Dying cells, including those originating from tumors, shed their DNA into the bloodstream and contribute to a pool of circulating fragments called cell-free DNA (cfDNA). Identifying the tissue origin of these DNA fragments from their epigenetic features has implications in various clinical contexts. Open chromatin regions (OCRs) are important epigenetic features of DNA that reflect cell types of origin. Profiling these features by DNase-seq, ATAC-seq, and histone ChIP-seq provides insights into tissue-specific and disease-specific regulatory mechanisms. Integration of genomic and epigenomic features for cancer detection by liquid biopsy has previously been reported. However, many multimodal analyses require large amounts of cfDNA input and/or multiple types of experiments to cover the genomic and epigenomic aspects of a single sample which is cost and time prohibitive. Thus, methods that capture genomic and epigenomic profiles in a single experiment type with low input requirements are of importance. Predicting OCRs from whole genome sequencing (WGS) data is one such approach. A previous study led to a tool named OCRDetector with promising performance. However, this method has certain limitations such as need for high-depth cfDNA WGS, prior information about OCRs distribution, and considering multiple features. Here, we applied a correlation clustering algorithm to predict OCRs. We used local sequencing depth as input to our algorithm. Multiple processing steps were then applied as follows: count normalization, discrete Fourier transform conversion, graph construction, graph cut optimization by linear programming, and clustering. To validate the proposed method, we compared the output of our predictions (OCR vs. non-OCR) with previously validated open chromatin regions related to human blood samples of the ATAC-db. The percentage of overlap between them is greater than 67%. Given that OCRs are mostly located in the transcription start sites (TSS) of expressed genes, we next compared the concordance between the predicted OCRs and TSS regions of housekeeping genes of human genome obtained from refTSS. We observed 70% of these regions among predicted OCRs, further confirming the validity of our approach. Overall, our proposed method which is based on unsupervised learning resulted in faster performance and decent accuracy and indicates predicting epigenomic features from cfDNA sequencing data is a promising and feasible alternative to approaches that are limited by low level of cfDNA in clinical samples.

**Introduction**

Open chromatin regions (OCRs) are nucleosome-depleted regions that can be bound by protein factors and play various roles in DNA replication, nuclear organization, and gene transcription [1]. Accordingly, they are suitable biomarkers for epigenomic analysis during disease progression particularly in cancer initiation and progression [2]. In the area of cancer early detection, especially for the inaccessible human tissues, such as solid tumors non-invasively monitoring the genomics and dynamics of their epigenomes is still largely unexplored. Circulating cell-free DNA (cfDNA) in the peripheral blood offers a promising and non-invasive approach to monitoring the genome and epigenome dynamics of the tissues [3]. However, most of the current cfDNA WGS studies only focus on the genetic aberrations in the cfDNA, such as SNPs or copy number variations (CNVs) and identifying epigenomic features from WGS data has not been considered that much.

There have been several experimental methods for epigenomic analysis like genome methylation analysis. Besides DNA methylation, several studies demonstrated cfDNA fragmentation pattern as a new epigenomic feature due to the meaningful nucleosomal distribution throughout the genome of cfDNA. Oberhofer et al showed that cell types can be inferred from genome nucleosome positioning data [4]. By considering this idea, Snyder et al tried to construct the nucleosomal footprints of cfDNA which resulted in identifying the tissue origin of cfDNA molecules [5]. These studies have been pursued by Sun et al to propose more accurate features for identifying the tissue origin of cfDNA as they validated the role of sequencing coverage imbalance and differentially phased fragment end signals [6].

The correlation between nucleosomal pattern and gene expression level is the main idea for the application of fragmentomics in cell identity. Cell identity is mostly carried out by gene transcription programs, which are determined by regulatory elements encoded in DNA. These particular regions which correspond to nucleosome-depleted regions, are called open chromatin regions (OCRs). Accordingly, during the cfDNA sequencing, DNA originating from the OCRs will be divided into small pieces due to the lack of histone protections and consequently lead to low depth of coverage in open chromatin regions. This feature is high of importance for further epigenomic studies [7].

There have been several methods for chromatin accessibility assays that most of the current methods separate the genome by enzymatic or chemical means and isolate either the accessible or protected locations. ChIP-Seq, DNase-seq, ATAC-seq, MNase-seq are among the well-known ones that use specific enzymes, such as Tn5 transposase, to process tissues. However, according to the previous studies, detecting OCRs in cfDNA related data based on these distinct approaches is challenging and highly time and cost consuming. To reduce these restrictions, researches have focused on exploring OCRs in the whole genome sequencing data so they can extract genomic and epigenomic features from the single cfDNA file. In this regard, Snyder et al used the fragment patterns of cfDNA to propose OCR detection method using window protection score (WPS), which is calculated in a 120 bp sliding window [5]. One of the premier studies in this area is the study by Ulz et al. They could infer accessibility of TF binding sites from cell-free DNA fragmentation patterns [8]. The most relevant study, named OCRDetector, calculates the coverage of cfDNA and the waveform of WPS for each 20000 bp interval from high coverage sequencing data[9]. As it is evident, there are few researches in this area and more accurate methods need to be developed for this aim. So, proposing a well-defined in silico analysis method for determining the genomic and epigenomic aspects of a cfDNA from a healthy or cancer individual by obtaining a single WGS data is of the interest. By identifying disease-specific or tissue-specific OCRs, it is possible to study the epigenetic mechanisms of cancers, predict potential markers, and analyze the tumor heterogeneity and subtyping.

As mentioned earlier, read depth is one of the preliminary features for investigating the OCRs in genomic data. As it has been reported, the read depth in the genomic regions related to OCRs shows reduction and the depth count fluctuation throughout the genome contains meaningful information that needs to be decoded in order to discover the possible OCRs. In the current study, we aimed to exploit the computational models comprising graph-based model for graph signal processing along with correlation clustering to predict the OCRs in cfDNA WGS data. The schematic overview of the proposed pipeline was illustrated in figure1.

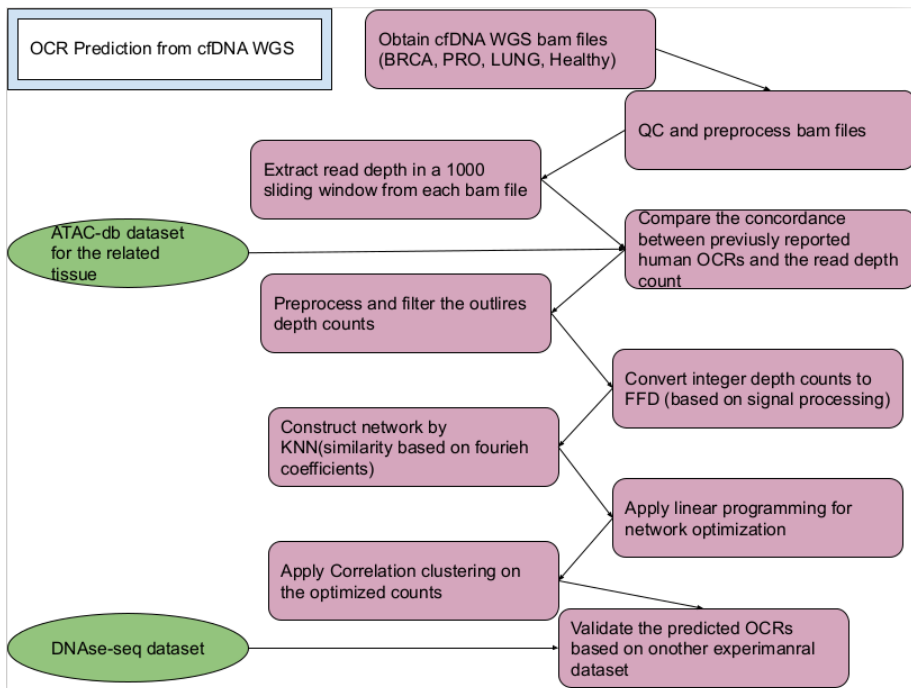

Figure 1: Schematic overview of OCR prediction method

**Materials and Methods**

In the current study, we aimed to predict reliable OCRs from cfDNA WGS data. Healthy and cancer related samples with low and high depth of sequencing were used in the development and validation of the proposed method.

**Data Collection**

**cfDNA WGS Data**

The cfDNA WGS dataset including healthy and cancer cfDNA samples with distinct sequencing depth was obtained with accession number of GSE71378 and PRJNA795275[5]. The mentioned raw sequencing data was aligned against hg19 with BWA aln and the obtained bam files with their indexes were used for further pre-processing steps. The detail of the input dataset was illustrated in Table 1.

**Chromatin Accessibility Data**

Along with the input sample files, previously reported and experimentally validated OCRs of human genome were obtained from ATACdb and the ENCODE Project for using as a gold standard data in our model evaluation. This database includes the OCRs from different tissue types. We collected the OCRs of some specific tissues that have the greatest contribution in the circulating blood to mimic the cfDNA content in healthy samples. These tissues were blood, bone marrow, liver, lung, and brain. The constructed OCR reference comprises 15 cell types with overall size of 1148576.

Table 1: The source of the data

| Data Type | Tissue/Primary cell type | Database | Accession Number |
|---|---|---|---|
| **cfDNA WGS Data** | Healthy | NCBI/GEO | GSE71378 |
|  | Healthy | NCBI/GEO | PRJNA795275 |
|  | Lung cancer | NCBI/GEO | GSM1833253 |
| **Chromatin Accessibility Data** | Healthy blood | ATACdb | http://www.licpathway.net/ATACdb/Download.php |
| | Healthy Bone Marrow | ATACdb | http://www.licpathway.net/ATACdb/Download.php |
| | Healthy Brain | ATACdb | http://www.licpathway.net/ATACdb/Download.php |
| | Healthy liver | ENCODE | https://www.encodeproject.org/search/?type=File&searchTerm=atac-seq&file_format=bed&biosample_ontology.term_name=liver&biosample_ontology.organ_slims=liver |
| | Healthy lung | ENCODE | |

**Pre-processing**

After obtaining the bam files of cfDNA, the regions belonging to the ENCODE Blacklist were omitted then the implemented in house python script was applied on the filtered cfDNA file to count the unique read depth in a 1000 bp sliding window. Choosing the sliding window with size of 1000 is because the average length of ATACdb reported OCRs is around 560 bps. The regions with less than 100 counts for their depth were omitted.

**The normalization and conversion of sequencing depth**

Next generation sequencing (NGS) approaches have been optimized to provide comparable technical performance for genomic and epigenomic analysis. However, it is possible to get biased results due to the technical errors during the sequencing process. Sequencing depth is one the important features in the area of liquid biopsy analysis. But the mentioned error needed to be targeted for the application of read depth in any in silico analysis. In this regard, we exploit the signal processing algorithms to convert the obtained depth values in each interval to a continuous data to cover the depth fluctuations of intervals that can be missed when working with single depth value. Here, the calculated depth counts were considered as discrete signals and then were converted to discrete-frequency representation by applying DFT algorithm. The Fourier transform converts a time domain representation of a signal into a frequency domain representation.

The Discrete Fourier Transform (DFT) of a sequence $\{s_m\}$ of $N$ numbers:

$s_0, s_1, ..., s_{N-1}$

The discrete Fourier transform of $\{s_m\}$ is a sequence $\{S_n\}$ of $N$ numbers:

$S_0, S_1, ..., S_{N-1}$

defined by the equation:s

$$S_n = \sum_{m=0}^{N-1} s_m w^{-mn}$$

$n=0,1,2,...,N-1$

where :

$$w = e^{j2\pi/N}$$

The parameters in the above formula were inferred from the real cfDNA data. As, we assumed each 100 data points represented a node and the sampling frequency is 200. Accordingly, the Nyquist limit was assigned as 100.

Then, the proposed algorithm used DFT coefficient to select the meaningful features.

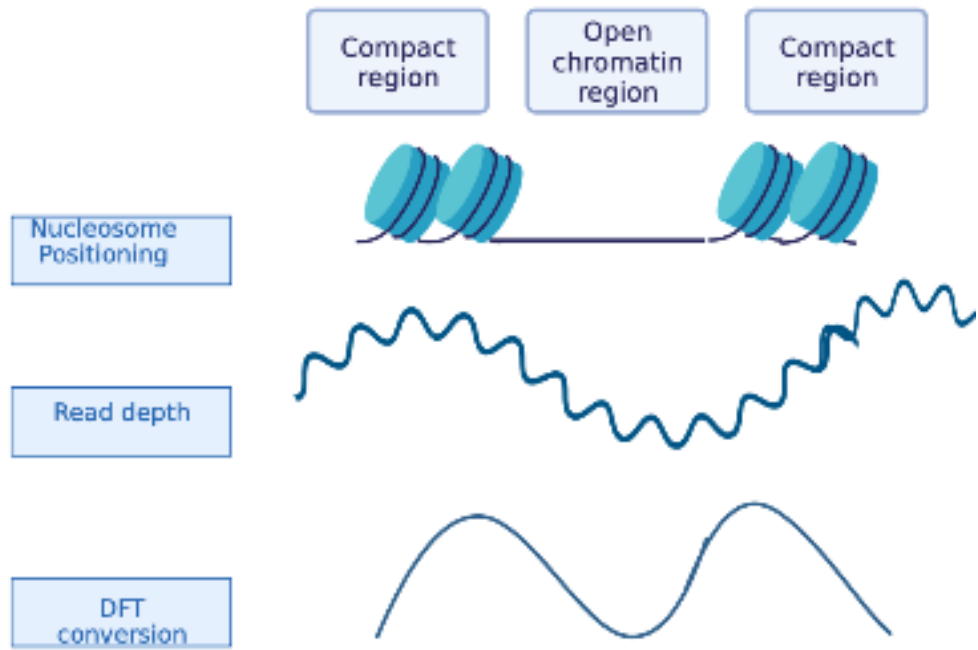

Figure 2: Schematic overview of OCR depth pattern

**Graph construction**

The biological related elements in the cell are usually connected to each other and work interactively. To model these interactions, networks play an important role. These networks can be useful to infer higher-order arrangements in the complex biological system that had been introduced as community detection. By considering this idea, we tried to apply the network-based attitude for OCR identification based on the similarity of their mathematical features that are introduced as Fourier coefficients in this study. Now, each extracted genome interval that is represented by its Fourier coefficients in the format of a vector considered as scattered points that needed to be connected. Here, using KNN's distance calculation method as an unsupervised learning method helped us make the graph. Euclidean distance was calculated as a distance metric between two points. Here, K with the number of 10 was implemented. Now, each node of the constructed graph represents the DFT and its features are DFT coefficients which describe the sinusoids that, when added, yield the original signal.

**Applying Correlation Clustering**

Now, an undirected graph with no weights is constructed. Even, in this step, we were able to label the nodes and infer the possible OCRs. However, this approach could result in huge amount of false positive data. To overcome this issue, we need to apply the optimization method to find the most similar communities. There have been several graphs cut algorithms to identify the similar clusters in a single graph. One of these methods is related to correlation clustering which uses linear programming to optimize the weighted graph in order to detect the similar communities. In this regard, the unweighted graph from the previous step, was converted to weighted network by assigning the distance score between each pair of nodes as a weight for the correspondent edge. Now, by obtaining the weighted graph, each edge should be labeled as positive weighted or negative weighted. To fulfill this task, the median of the distance scores was measured and the edges with greater weight of the mentioned median were considered as positive and the rest as negative. By obtaining this data, the algorithm proposed by Charikar et al was implemented [10]. This algorithm bounds the number of misclassified edges incident on any node that referred as correlation clustering with local guarantees. As described in the original article, the Min Max disagreements on general weighted graph was carried out by first computing the linear programming (LP) metric and then performing a layered clustering. The objective function for Min Max disagreements on general weighted graph was implemented.

**The Local Minimization of Disagreements**

During the relaxation process a metric d will be considered on each node of the constructed graph. Now for each node u, a parameter entitled D(u) relating to the total fractional disagreement of edges incident on it will be calculated. The objective function of the minimization phase is[10]:

$$\text{Min} \left( \sum_{v:(u,v)\in E^+} c_{u,v} d(u,v) + \sum_{v:(u,v)\in E^-} c_{u,v}(1 - d(u,v)) \right)$$

**Local Maximization of Agreements**

In this step, the minimum total weight of correctly classified edges incident on any node u is maximized. For example, given a set of partitions as S that for u ∈ Si:

$$agree_S(u) \triangleq \sum_{v \in Si:(u,v)\in E^+} c_{u,v} + \sum_{v \notin Si:(u,v)\in E^-} c_{u,v}$$

At the end by considering both of the above-mentioned optimizations, the final objective function is generated and solved by LP.

This optimization was solved by SciPy package. The output of the LP was the list of pair nodes with their optimized weight value that needs to be rounded in the next step. After calculating the optimized values of the weights, the layered clustering was implemented. The layered clustering was imposed on the Here, the number of clusters can be pre-identified and at the end five clusters were generated with optimum distance metrics.

**OCR identification**

After obtaining the clusters, we need to evaluate the model performance and annotate the biological aspect of the results. Therefore, in this phase the probable OCRs in cfDNA samples will be determined. Each cluster components were extracted and the correspondent regions were compared against the validated human OCRs obtained from ATACdb and ENCODE. The regions that overlap between these two datasets along with the distance of absolute value of 500 bps around the validated regions were reported as candidate OCRs.

**Model validation**

The validation of the model should be carried out in two resolutions:

1- calculating the mathematical metrics related to graph cut algorithm, and

2- the biological interpretation of the model results. As mentioned in the previous step, the candidate OCRs were investigated based on the real validated human OCRs. The proposed model was applied on the other cfDNA samples related to a cancer dataset (lung cancer) to further assess its performance.

There have been several types of experiments to find the OCRs like ATAC-Seq, DNase-seq, and etc. Here, another publicly available database of DNase-seq (http://www.uwencode.org/)was considered as well for comparative analysis.

## Results

We aimed to propose a computational method for OCR prediction that can handle most of the technical and experimental variations. Accordingly, the healthy and cancerous samples with distinct sequencing depth were obtained.

The depth distribution in 1000 bps bins of the IH02(cfDNA healthy sample) with average sequencing depth of 30X is illustrated in Figure 3. 2891984 regions remained after removing the bins with less than 100 counts.

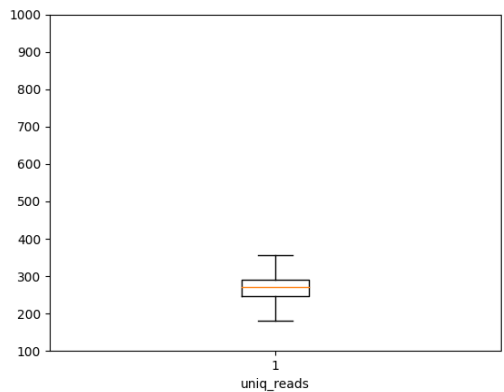

Figure 3: Depth distribution in 1000 bps bins with average sequencing depth in the cfDNA sample

The OCR size distribution in validated database related to the healthy blood data is:

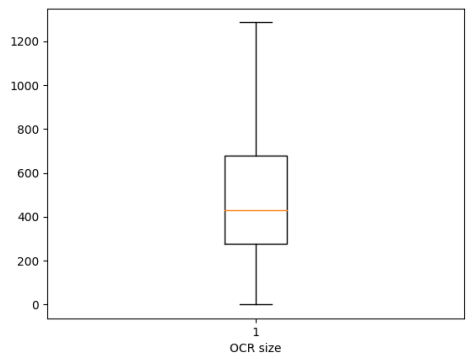

Figure 4: The OCR size distribution in validated database (ATACdb)

As it is illustrated in the below figure5, the raw depth distribution varies significantly between the validated OCRs and the extracted candidate regions from cfDNA samples. However, after converting the depth feature to signal by DFT method the concordance between their distribution pattern throughout the genome increases.

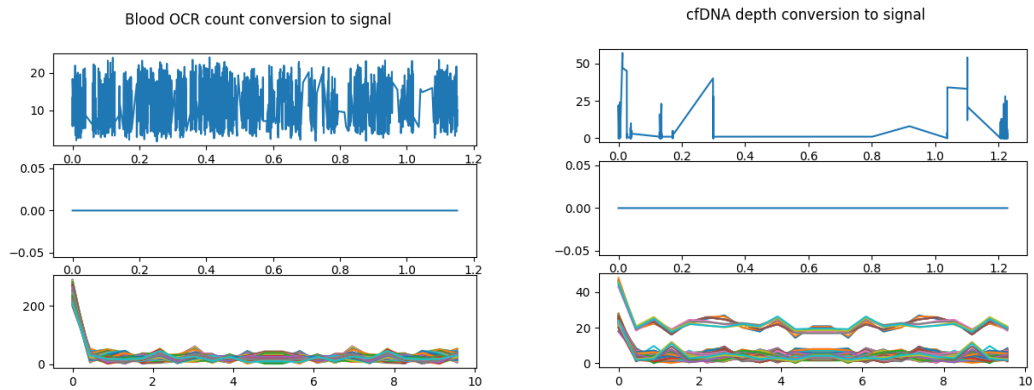

Figure 5: Comparison of extracted genomic regions with ATACdb reported regions

By achieving decent concordance between the extracted and experimentally validated regions, the Fourier coefficients were used as the selected features for further analysis and constructing the graph. The constructed network consists of 68575 nodes.

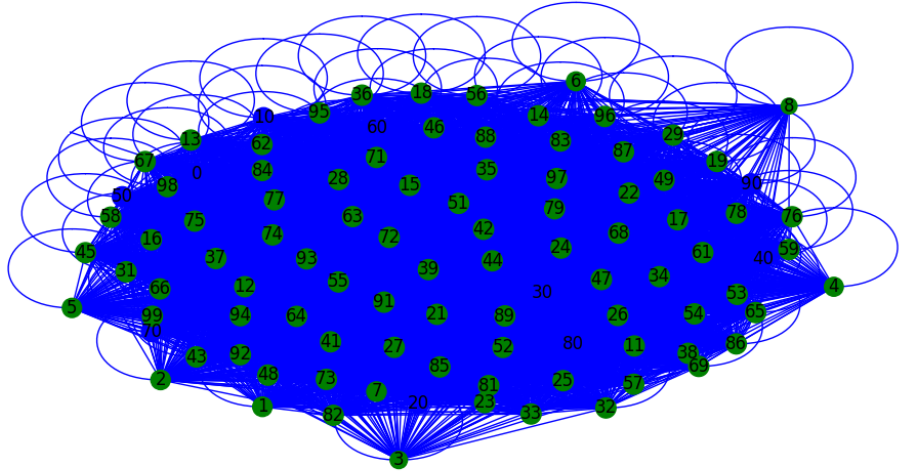

Figure 6: Constructed graph by KNN(Schematic overview)

After optimizing and rounding the weights based on the charikar's algorithm, the prepared matrix was fed to the layered clustering algorithm. By this approach the number of clusters can be determined by the user. Choosing the optimum number of clusters can be done by maximizing the number of similar edges in each cluster along with dissimilar edges between the cluster. Here, after checking the clustering errors five clusters were generated, four of them are connected to each other which showed the similarity between these distinct clusters.

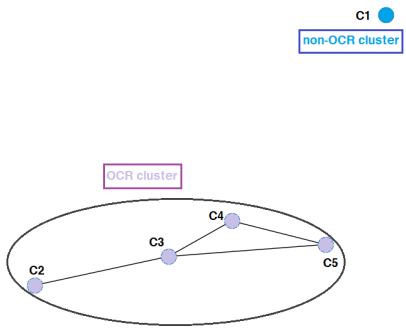

Figure 7: Cluster representation of constructed graph

By this algorithm we could obtain five clusters, one completely departed and the remained four connected to each other with containing most of the possible OCRs when comparing with the reference human OCRs. The number of possible OCRs in the detached cluster was less than 4% and we can call it the non-OCR cluster. And the remained four connected clusters as OCR cluster with more than 67% overlap with validated OCRs.

The length distribution of the experimentally validated human genome OCRs was illustrated in above figure3. The sizes of the OCRs related to blood provided by ATACdb were 276, 432, and 680 for Q1, Q2, and Q3 quantile respectively.

In the other hand, the performance of the proposed method was assessed on a specific region annotated as TSS regions of housekeeping genes. Naturally, these regions include several numbers of OCRs and we expected to identify more accurate regions as OCRs in the cfDNA samples. As it is illustrated in Figure 7 among the four connected clusters that represent 36629 cfDNA-derived nodes around 52% were overlapped with the human genome TSS collected in refTSS database. And about 78% of them were related to the human housekeeping genes that normally have higher gene expression level and accordingly their OCRs are dominant.

As mentioned earlier the healthy blood OCRs were obtained from two databases. This data gathers the OCRs of different blood cells such as CD56Bright natural killer cell, Monocyte, Natural Killer cell, Cultured T cell, CD8 T cell, CD19+ B cell, CD34+ HSPC, and CD4+ T cell. We know that one of the main application of OCRs as an epigenomic feature is to identify the tissue(cell) origin of the samples. So, the reference of human genome OCRs related to several tissue types was collected to mimic the cfDNA content. This reference includes the reported OCRs from blood, bone marrow, liver, lung, and brain. This locally-built reference comprises    unique OCRs from several tissue types. The comparative analysis between the predicted OCRs and the validated ones has been done and the result could satisfy the expectations in the distribution of different tissue's OCRs.

As illustrated in below figure, blood related OCRs occupy the first place which is in agreement with the cfDNA content.

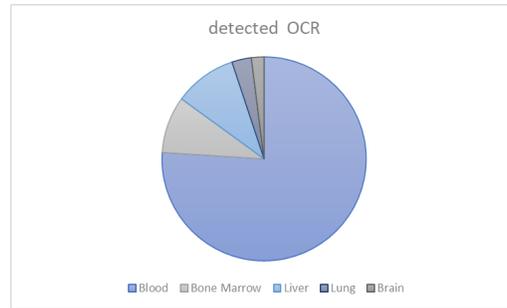

Figure 8: Tissue contribution of OCRs

**Assessing non-OCR cluster**

The genome is constructed of euchromatin and heterochromatin regions that differs based on the compaction of the genome as the euchromatin part has less compact structure and high possibility of OCR occurrence. In the other side, the heterochromatin part is highly condensed with very low number of genes which reduces the occurrence of OCRs in them. Accordingly, it is assumed that most of the non-OCRs should belong to the heterochromatin part of the genome. Hopefully, there are several signatures that can be detected in the heterochromatin regions such as the repeated regions. These regions mainly comprise of Microsatellites, Minisatellites, Variable number tandem repeats (VNTR), short interspersed nuclear element (SINE), long interspersed nuclear element (LINE), LTR retroposons, and DNA transposons.

As mentioned in the previous section, the meaningful amount of the data was identified as non-OCRs. Here, to evaluate their nature, the proposed non-OCRs were compared against the locally constructed reference for human genome repeated regions. To make this reference, annotated regions from several databases were collected. The table below shows the obtained data.

**The source of the data**

| Reference | Data Type | Database |
|---|---|---|
| **Human genome repeated regions** | MicroSatellite | MicroSatellite DataBase (MSDB) |
| | All reapeted regions | UCSC Repeat Browser |
| | RETROTRANSPOSONS | dbRIP |

This reference includes 5585519 unique regions with different length size. Now, the regions belong to the non-OCR cluster(C1) were extracted and mapped to the repeated regions reference file to identify the percentage of overlap between them.

**Model evaluation on cancer data**

Further evaluations have been conducted to deeply investigate the proposed model performance. Along with the healthy samples, the cancer cfDNA sample was also obtained for analysis. Here, a high quality cfDNA WGS sample of a lung cancer (IC20) with accession number of GSM1833245 was used. This data obtained from a Lung cancer with coverage of 23.38. We choose this dataset due to its high depth of sequencing along with the significant contribution of lung tissue in cfDNA content. Similar to the previous sections read counts in a sliding window of 1000bps were extracted and converted to discrete signal by DFT. Here, after applying the correlation clustering three distinct clusters were generated with the one containing the most overlaps with the validated OCRs. For the comparison, the previously mentioned locally developed OCR reference was enriched with the lung cancer OCRs obtained from ATACdb.

Along with the lung cancer data, liver cancer analysis has been considered as well. The high quality cfDNA WGS data related to a liver cancer (Hepatocellular carcinoma) was obtained under the accession number of GSM1833242 with coverage of 42.08. meanwhile the validated OCRs of a liver cancer was obtained from CATA database which comprises 121607regions for liver cancer after processing. Like the previous sections the clustering and the comparative analysis to label the cluster's components were carried out.

**The performance of layered clustering**

In the current study, the LP method proposed by Charikar et al was implemented to optimize the graph cut algorithm. However, there have been several other optimizations approaches for correlation clustering.

These algorithms differ mainly based on their solving method of the objective function and rounding phase.

Here, the results of our method were compared with two other optimization algorithms proposed by Demaine et al and Ailon et al.

We observed that all these algorithms have positive impact on reducing the number of false positives. And our method had better performance on the speed factor due to measuring the local optima instead of global one.

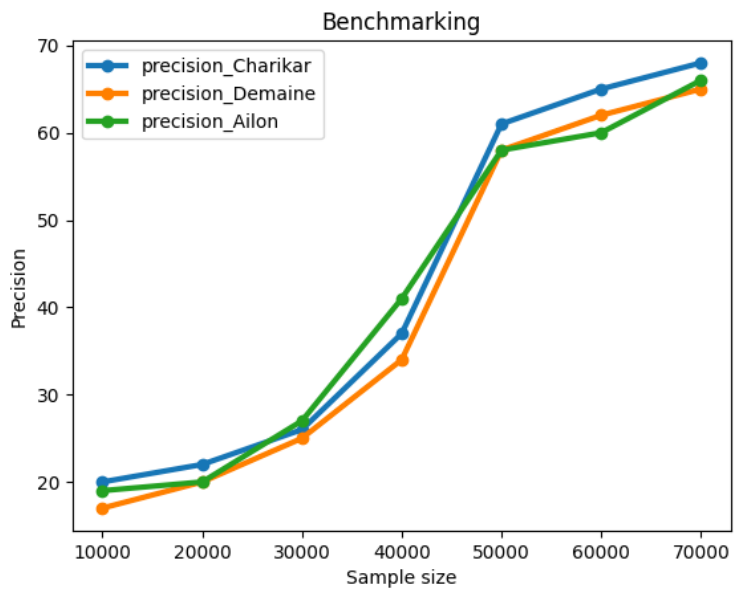

Figure 9: Comparing several optimization methods

**Discussion**

In the current study, we aimed to propose a computational method for OCR detection form cfDNA WGS data based on sequencing depth due to the high expense and experimental complexity of ChIP-Seq, DNase-seq, ATAC-seq, and MNase-seq. OCRs are one of the most important epigenetic features that can be investigated for cancer diagnosis and as it is evident cfDNA analysis plays significant role in cancer early detection and monitoring. Because, cfDNA data has its own unique variations when comparing with tissue-derived NGS data, its application in clinic needs integration of genetic and epigenetic features to achieve more accurate results. However, including several lab experiments to cover genetic and epigenetic aspects of the samples requires complex in silico(bioinformatic) analysis along with wet lab costs and labor. By considering the mentioned restrictions and obstacles, providing a bioinformatic method for investigating genetic and epigenetic features from the single cDNA WGS data is of special interest and importance. In this study, by focusing on the sequencing depth values instead of integrating several other features along with exploiting fundamental computational approaches such the signal processing and graph correlation clustering, we tried to implement OCR predictive model. Also, we have analyzed the relative contributions of various cells in plasma DNA. The diagnostic potential of this method was assessed in cancer diseases as well. As mentioned in the previous section, the OCRs are distributed throughout the genome with most frequency in the promoter, transcription factor binding site, and enhancer. The results obtained by our method, is in agreement with this fact and the overlap between the predicted OCRs and these three types of regions was more significant.

Having network-based attitude for OCR prediction along with paying significant attention on signal related mathematical features for read depth count was done for primer time in the current study and this approach may help us to reduce the technical biases originated from sequencing depth. The accuracy of the proposed method has been evaluated by using several experimentally proved databases and there is a decent positive correlation between them.

As we know, cfDNA WGS data harbors different types of variations that can have negative impact on extracting any accurate feature and result from it. Existence of black-list regions in NGS data that lead to anomalous, unstructured, or high signal independent of cell line or experiment is the preliminary issue that was targeted in this study by removing them from the original bam files. The other problem causing factor is related to the chromosomal alterations such as translocation, indels. So, by applying the continuous counts instead of discrete ones and focus on the depth fluctuations throughout the genome we could reduce the number of false positive to some extent. Meanwhile, the fast Fourier transform (FFT) that is an efficient algorithm to compute the discrete Fourier transform (DFT) and its inverse can be used for detecting repeats over the genome because its analysis periodicity in data. These repeated regions that are mostly categorized as Microsatellites, Minisatellites, Variable number tandem repeats (VNTR), short interspersed nuclear element (SINE), long interspersed nuclear element (LINE), LTR retroposons, and DNA transposons are able to influence the raw depth counts and result in biased findings. Therefore, FFT can help us to detect and remove them to reduce the errors and achieve reasonable results. There are several other challenges that need to be considered to increase the accuracy and performance of our model.

Our proposed method is following unsupervised manner and to predict the possible OCRs, we do not need to insert any prior knowledge to the predictive model and only based on the similarity metrics in the context of graph algorithms we could predict probable OCRs with decent precision. It is worth mentioning that at first phase the graph was generated by KNN in unsupervised mode that by itself could be used to predict the OCRs. However, to reduce the number of false positives the correlation clustering was applied in the next phase to maximize the agreements between similar nodes along with disagreement between dissimilar ones. However, there are several biological factors that influence the sequencing depth and need to be considered such as which will be added to the original version of the method in future works.

## Data Availability

All the raw cfDNA WGS samples were obtained from NCBI GEO under the accession number of GSE71378 and GSE114511.

## Conflicts of Interest

The authors declare no conflict of interest.

## References

[1] Robert E Thurman, Eric Rynes, Richard Humbert, Jeff Vierstra, Matthew T Maurano, Eric Haugen, Nathan C Sheffield, Andrew B Stergachis, Hao Wang, Benjamin Vernot, et al. The accessible chromatin landscape of the human genome. Nature, 489(7414):75–82, 2012.

[2] William A Flavahan, Elizabeth Gaskell, and Bradley E Bernstein. Epigenetic plasticity and the hallmarks of cancer. Science, 357(6348):eaal2380, 2017.

[3] Abel Jacobus Bronkhorst, Vida Ungerer, and Stefan Holdenrieder. The emerging role of cell-free dna as a molecular marker for cancer management. Biomolecular detection and quantification, 17:100087, 2019.

[4] Oberhofer, Angela, et al. "Tracing the Origin of Cell-Free DNA Molecules through Tissue-Specific Epigenetic Signatures." Diagnostics 12.8 (2022): 1834.

[5] Snyder, Matthew W., et al. "Cell-free DNA comprises an in vivo nucleosome footprint that informs its tissues-of-origin." Cell 164.1-2 (2016): 57-68.

[6] Sun, Kun, et al. "Orientation-aware plasma cell-free DNA fragmentation analysis in open chromatin regions informs tissue of origin." Genome research 29.3 (2019): 418-427.

[7] Anirudh Natarajan, Galip Gürkan Yardımcı, Nathan C Sheffield, Gregory E Crawford, and Uwe Ohler. Predicting cell-type–specific gene expression from regions of open chromatin. Genome research, 22(9):1711–1722, 2012.

[8] Ulz, Peter, et al. "Inference of transcription factor binding from cell-free DNA enables tumor subtype prediction and early detection." Nature communications 10.1 (2019): 1-11.

[9]Wang, Jiayin, et al. "OCRDetector: Accurately Detecting Open Chromatin Regions via Plasma Cell-Free DNA Sequencing Data." International Journal of Molecular Sciences 22.11 (2021): 5802.

[10] Charikar, Moses, Neha Gupta, and Roy Schwartz. "Local guarantees in graph cuts and clustering." International Conference on Integer Programming and Combinatorial Optimization. Springer, Cham, 2017.